\documentclass[12pt]{article}
\newcommand{\noi}{\noindent}
\newcommand{\eq}{\begin{equation}}
\newcommand{\en}{\end{equation}}
\newcommand{\eqa}{\begin{eqnarray}}
\newcommand{\ena}{\end{eqnarray}}

\newcommand{\oeta}{{\overline \eta}}

\newcommand{\opsi}{{\overline \psi}}

\newcommand{\ozeta}{{\overline \zeta}}

\newcommand{\cam}{{\cal M}}
\newcommand{\cao}{{\cal O}}
\newcommand{\capbig}{{\cal P}}
\newcommand{\car}{{\cal R}}

\newcommand{\hatJ}{{\widehat J}}
\newcommand{\hatH}{{\widehat H}}

\newcommand{\hatN}{{\widehat N}}
\newcommand{\hatO}{{\widehat O}}

\newcommand{\hatT}{{\hat T}}

\newcommand{\hatV}{{\widehat V}}

\newcommand{\bfB}{{\bf B}}
\newcommand{\bfC}{{\bf C}}

\newcommand{\tr}{\mbox{Tr}\,}
\newcommand{\re}{\mbox{Re}\,}

\newcommand{\oq}{{\overline q}}

\newcommand{\vx}{{\vec x}}
\newcommand{\vy}{{\vec y}}
\newcommand{\vz}{{\vec z}}

\newcommand{\spr}{s^{\prime}}

\newcommand{\lra}{\longrightarrow}
\begin{document}

\renewcommand{\baselinestretch}{1.1}
\renewcommand{\thefootnote}{\arabic{footnote}}
\setcounter{footnote}{0}
\renewcommand{\theequation}{\arabic{section}.\arabic{equation}}
\renewcommand{\thesection}{\arabic{section}}
\language0

%%%%%%%%%%%%%%%%%%%%%%%%%%%%%%%%%%%%%%%%%%%%%%%%%%%%%%%%%%%%%%%%%%%%%%%
%  \thispagestyle{empty}
%  \pagebreak

%  \hbox{~}
%  \pagenumbering{roman}
%  \setcounter{page}{1}
%  \tableofcontents
%  \pagebreak
%  \pagenumbering{arabic}
%  \setcounter{page}{1}
%%%%%%%%%%%%%%%%%%%%%%%%%%%%%%%%%%%%%%%%%%%%%%%%%%%%%%%%%%%%%%%%%%%%%%%
%  \renewcommand{\thefootnote}{\fnsymbol{footnote}}
%  \setcounter{footnote}{1}
%%%%%%%%%%%%%%%%%%%%%%%%%%%%%%%%%%%%%%%%%%%%%%%%%%%%%%%%%%%%%%%%%%%%%%%

\hbox{}
\noindent ~~~June  2002

\vspace{1.0cm}
\begin{center}

{\Large Transfer matrix and nonperturbative renormalization of
fermionic currents in lattice QCD}

\vspace*{1.0cm}
{\large V.K.~Mitrjushkin$\mbox{}^{a,b}$}

\end{center}

\vspace*{0.5cm}

{\sl \noindent
 \hspace*{12mm} $^a$ Joint Institute for Nuclear Research, 141980 Dubna,
          Russia \\
 \hspace*{12mm} $^b$ Institute of Theoretical and Experimental Physics,
                    Moscow, Russia}

\vspace{10mm}
\begin{abstract}

The functional integral representation for fermionic observables on
the lattice is studied. In particular, Grassmannian representations of
the scalar $\hatJ^{(S)}$ and pseudoscalar $\hatJ^{(P)}$ currents and
pseudoscalar correlator are derived. It is also discussed the
connection between the fermionic Fock space and boundary conditions
along the time direction.

\end{abstract}

\section{Introduction}\setcounter{equation}{0}

The `functional' integral approach to the quantization with lattice
regularization in the euclidean space is very convenient for numerical
calculations and gives a possibility for the non--perturbative study of
the gauge theories \cite{wil1}. However, a consistent quantization
scheme needs a canonical Hamiltonian (or transfer matrix) approach to
answer questions that do not obviously hold in the functional integral
formulation.

The canonical quantization formalism is necessary to define boundary
conditions for $~U_{x\mu}~$ and $~\psi,\opsi_x~$ in the integrals in
eq.'s ~(\ref{partition1}),(\ref{obs1}), in particular along the forth
(imaginary time) direction. It is the canonical quantization formalism
which establishes the connection between correlators of currents (e.g.,
pseudoscalar current $\hatJ^{(P)}$) and masses of corresponding
particles.

The standard Wilson action $S_W$ with $SU(N_c)$ gauge group is
\cite{wil1}

\eq
S_W(U;\psi;\opsi) = S_G(U) + S_F(U;\psi;\opsi)~,
\en

\noi where pure gauge part $S_G$ and fermionic part $S_F$ are given by

\eqa
S_G &=& \beta \sum_x\sum_{\mu >\nu}
\left[ 1 -\frac{1}{N_c} \re\tr \Bigl(U_{x\mu}U_{x+\mu;\nu}
U_{x+\nu;\mu}^{\dagger}U_{x\nu}^{\dagger} \Bigr)\right]~;
\qquad \beta =\frac{2N_c}{g^2}~;
\nonumber \\
\nonumber \\
S_{F} &=& \sum_{f=1}^{N_f}\sum_{x,y} \sum_{s,\spr=1}^{4}
\opsi_{x}^{f,s} \cam^{f;s\spr}_{xy} \psi_{y}^{f,\spr}
\equiv \opsi\cam\psi~.
\ena

\noi In equations above $~U_{x\mu}\in SU(N_c)~$ are gauge fields,
$~\psi_x,\opsi_x~$ are fermionic Grassmann variables and $g=g_0(a)$ is
a bare coupling, $~a$ being a lattice spacing. Wilson's fermionic
matrix $~\cam(U)~$ (with $N_f=1$) is given by

\eq
\cam_{xy} = \delta_{xy} - 2\kappa\sum_{\mu}\Bigl[ \delta_{y, x+\mu}\,
P^{(-)}_{\mu}U_{x\mu} + \delta_{y,x-\mu}\, P^{(+)}_{\mu}
U_{x-\mu,\mu}^{\dagger} \Bigr]~,
\en

\noi where $\kappa$ is a hopping parameter and
$~P^{(\pm)}_{\mu} = (1\pm\gamma_{\mu})/2~$.
The action $S_W$ is invariant with respect to the local
transformations

\eqa
U_{x\mu} &\stackrel{\Omega}{\lra}& U_{x\mu}^{\Omega}
= \Omega_x U_{x\mu}\Omega_{x+\mu}^{\dagger}~;
\qquad
\psi_x \stackrel{\Omega}{\lra} \Omega_x\psi_x~;
\qquad
\opsi_x \stackrel{\Omega}{\lra} \opsi_x\Omega_x^{\dagger}~,
\ena

\noi with $~\Omega_x \in SU(N_c)~$. The partition function $Z_W$ is

\eq
Z_W = \int\! [dU][d\psi d\opsi]~e^{-S_W(U;\psi;\opsi)}~,
         \label{partition1}
\en

\noi where $~[dU]~$ denote product $~\prod_{x\mu}dU_{x\mu}~$, etc..
The average of any functional $~\cao(U;\psi;\opsi)~$ is chosen to be

\eq
\langle \cao\rangle = \frac{1}{Z_W}\int\! [dU][d\psi d\opsi]
~\cao(U;\psi;\opsi)\cdot e^{-S_W(U;\psi;\opsi)}~.
            \label{obs1}
\en

\noi Given some boundary conditions, the average $~\langle \cao\rangle~$
is (mathematically) well defined and can be calculated numerically.

In the canonical quantization approach, the connection between the
transfer matrix $~\hatV~$ and the Hamiltonian $~\hatH~$ is given by

\eq
\hatV = e^{-a\hatH}~,
\en

\noi The corresponding partition function $Z_H$ is

\eq
Z_H = \tr \left( e^{ -\frac{1}{T}\hatH } \right)_{colorless}
= \tr \left( \hatV^{N_4} \right)_{colorless}~,
         \label{partition2}
\en

\noi where $~N_4~$ is the lattice size along the forth direction,
$~T=1/aN_4 = 1/L_4~$ and the trace is defined on some colorless space
of states. The consistency between two definitions of the partition
function given in eq.~(\ref{partition1}) and eq.~(\ref{partition2}),
i.e. $~Z_W=Z_H \equiv Z~$, defines the transfer matrix $~\hatV~$
\cite{cre1,lues,cre2} . The average of any field operator
$~{\widehat\cao}~$ is

\eq
\langle{\widehat\cao}\rangle = \frac{1}{Z}\tr \left( {\widehat\cao}
\hatV^{N_4} \right)_{colorless}~.
            \label{obs2}
\en

\noi Let $~|\Psi_k\rangle~$ be eigenstates of the transfer matrix
$\hatV~$ with eigenvalues $~\lambda_k~$

\eq
\hatV|\Psi_k\rangle = \lambda_k|\Psi_k\rangle~;
\qquad
\lambda_k = e^{-aE_k}~,
\en

\noi and $~E_k~$ are eigenvalues of the Hamiltonian.  Then

\eq
\langle {\widehat\cao}\rangle \sim \sum_{k\ge 0} e^{-\frac{1}{T}E_k}
\cdot \langle\Psi_k |{\widehat\cao}|\Psi_k\rangle~.
\en

\noi Similar expressions can be written for the correlators of
currents, e.g. pseudoscalar current $\hatJ^{(P)}$, etc..

The question of interest is the connection between operators
$\widehat\cao$ defined as normal products of creation and annihilation
operators and corresponding functionals $~\cao(U;\psi;\opsi)~$ in the
functional integral approach. Another problem of interest is the choice
of the boundary conditions along the imaginary time direction.

This paper is dedicated to the connection between the operator
(canonical quantization) formalism \cite{cre1,lues,cre2} and functional
integral approach \cite{wil1}. The transfer matrix formalism  is
briefly reviewed in the second section. In the third section the
statistical averages of the scalar $\hatJ^{(S)}$ and pseudoscalar
$\hatJ^{(P)}$ currents are calculated as well as pseudoscalar
correlator. Boundary conditions are discussed in the forth section. The
last section is reserved for conclusions.

\section{Transfer matrix formalism}
\setcounter{equation}{0}

Let us give an outline of the transfer matrix formalism following the
paper \cite{lues}. The main modification is connected with the
introduction of the projection operator $P_0$ (see below) which is
necessary to take into account the Gauss law \footnote{In paper
\cite{lues} the gauge $U_{x;4}=1$ has been chosen which is not possible
on the finite torus.}.

Let $c_i^{\dagger}(c_i)$ and $d_i^{\dagger}(d_i)$  be
creation(annihilation) operators of fermions and antifermions,
respectively, that satisfy canonical anticommutation relations :

\eq
\Bigl[ c_i,c^{\dagger}_j\Bigr]_{+} = \Bigl[ d_i,d^{\dagger}_j\Bigr]_{+}
= \delta_{ij}~,
\en

\noi and other anticommutation relations are equal to zero. Indices
$i,j$ are composite : $i=(\vx;\alpha;s)~$ where $\vx~$ is a three
dimensional coordinate, $\alpha=1,\ldots,N_c~$ is a color index and
$s=1,2~$ is a spin index. Therefore, $~i,j=1;\ldots;N~$ where
$~N=2N_cV_3~$ and $V_3$ is a threedimensional volume.

Let $~\{U_{\vx;k}\}~$ and $~\{U_{\vx;k}^{\prime}\}~$ be two
configurations of the gauge fields defined on the spacelike links
$~l_s=(\vx;k)~$. The transfer matrix $\hatV$ is an integral operator
with respect to the gauge fields. Its kernal $~V(U;U^{\prime})~$ is an
operator in the fermion Hilbert space and has the following form
\cite{lues}

\eqa
V(U;U^{\prime}) &=& \hatT^{\dagger}_F(U)V_G(U;U^{\prime})
\hatT_F(U^{\prime})~;
          \label{transf_luescher}
\\
\nonumber \\
\hatT_F(U) &=& C_0\cdot \exp\Bigl\{ d^TQ(U)c\Bigr\}
\exp\Bigl\{ -c^{\dagger}R(U)c - d^{\dagger}R^T(U)d\Bigr\}~;
\nonumber \\
\nonumber \\
\hatT^{\dagger}_F(U) &=& C_0 \cdot \exp\Bigl\{ -c^{\dagger}R(U)c
- d^{\dagger}R^T(U)d\Bigr\}\exp\Bigl\{ c^{\dagger}Q(U)
d^{\dagger\, T}\Bigr\}~,
\nonumber
\ena

\noi where $~V_G(U;U^{\prime}) \equiv \langle U|\hatV_G
|U^{\prime}\rangle~$ corresponds to a pure gauge part \cite{cre1}.
Hermitian matrices $~Q,R~$ are given by

\eqa
e^R &=& \frac{1}{\sqrt{2\kappa}}\, \bfB^{1/2}~;
\nonumber \\
\nonumber \\
\bfB_{\vx\vy} &=& \delta_{\vx\vy} - \kappa\sum_{k=1}^3\Bigl[
\delta_{\vy, \vx+k}U_{xk}+ \delta_{\vy,\vx-k}U_{x-k,k}^{\dagger} \Bigr]~,
\ena

\noi and

\eq
Q_{\vx\vy} = \frac{i}{2}\sum_{k=1}^3\Bigl[ \delta_{\vy, \vx+k}
U_{xk} - \delta_{\vy,\vx-k}U_{x-k,k}^{\dagger} \Bigr]\cdot \sigma_k~.
% ~\equiv~ i\sum_{k=1}^3 \bfC_{\vx\vy}^{(k)}(x_4) \cdot \sigma_k~.
\en

\noi $C_0 = C_0(U;U^{\prime})$ is a constant to be defined later and
$\sigma_k$ are Pauli matrices.

It is convenient to define Grassmannian coherent states

\eq
|\eta\zeta\rangle = \exp\left\{ \sum_{\vx} \Bigl(
c_{\vx}^{\dagger}\eta_{\vx} + d_{\vx}^{\dagger}\zeta_{\vx}\Bigr)\right\}
|0\rangle~;
\qquad
\langle\eta\zeta| = \langle 0|\exp\left\{ \sum_{\vx} \Bigl( \oeta_{\vx}
c_{\vx} + \ozeta_{\vx}d_{\vx} \Bigr)\right\}~,
\en

\noi where $~\eta_{\vx},\ldots,\ozeta_{\vx}~$ are some Grassmannian
variables (spin and color indices suppressed). It is easy to see that

\eq
c_{\vx}|\eta\zeta\rangle = \eta_{\vx}|\eta\zeta\rangle~;
\qquad
\langle\eta\zeta|c_{\vx}^{\dagger} = \langle\eta\zeta|\oeta_{\vx}~,
               \label{properies_coher}
\en

\noi and

\eq
\langle\eta\zeta|\eta^{\prime}\zeta^{\prime}\rangle
= e^{\oeta\eta^{\prime} + \ozeta\zeta^{\prime}}~,
\qquad
{\hat 1} = \int\! [d\oeta d\eta][d\ozeta d\zeta]
~e^{-\oeta\eta -\ozeta\zeta}|\eta\zeta\rangle
\langle\eta\zeta|~.
            \label{completeness}
\en

\noi To prove the equivalence between the transfer matrix and
functional integral approaches one can start with

\eq
Z_H = \tr \left( \hatV^{N_4} \right)_{colorless}
= \tr \left( \hatV^{N_4} P_0\right)~,
\en

\noi where $P_0$ is the projection operator on the colorless state

\eq
P_0 = \int\![d\Lambda]~\car(\Lambda)~;
\qquad
[d\Lambda] = \prod_{\vx}d\Lambda(\vx)~,
\en

\noi and $~\car(\Lambda)~$ is the gauge transformation operator.
In particular,

\eq
\car(\Lambda)\chi_{\vx}\car^{\dagger}(\Lambda)
= \Lambda^{\dagger}_{\vx}\chi_{\vx}~;
\qquad
\car(\Lambda)\chi_{\vx}^{\dagger}\car^{\dagger}(\Lambda)
= \chi_{\vx}^{\dagger}\Lambda_{\vx}~,
\en

\noi where

\eq
\chi_{\vx} = \left(
\begin{array}{c}
   c_{\vx}                \\
   d^{\dagger\, T}_{\vx}  \\
\end{array} \right)~;
\qquad
\chi_{\vx}^{\dagger} = ( c_{\vx}^{\dagger} ~~~d_{\vx}^T)~.
\en

\noi To obtain the functional integral representation of $Z_H$ one
needs to represent $~\hatV^{N_4}~$ as a product : $~\hatV^{N_4} =
\hatV\cdot \hatV \cdot \ldots \cdot\hatV~$ and insert unit operators
$~{\hat 1}~$ defined in eq.~(\ref{completeness}).
For every time slice $x_4$ one can introduce new Grassmannian
variables $~\psi_{\vx}(x_4),\opsi_{\vx}(x_4)$~:

\eq
\psi_{\vx}(x_4) = \sum_{\vy} \bfB^{-1/2}_{\vx\vy}\left(
\begin{array}{c}
   \eta_{\vy}                \\
   \ozeta^T_{\vy}  \\
\end{array} \right);
\qquad
\opsi_{\vx}(x_4) = \sum_{\vy} \Bigl(\oeta_{\vy} ~~~ -\zeta_{\vy}^T \Bigr)
\bfB^{-1/2}_{\vy\vx}~.
                  \label{new_grass}
\en

\noi Therefore, $~[d\oeta d\eta][d\ozeta d\zeta] =  J^2 \cdot [d\opsi
d\psi]~$. The constant $~C_0~$ is chosen to cancel the Jacobian $J$ of
this transformation.
Assuming that the fermionic Fock space spanned by all possible fermionic
states, one obtains

\eq
Z_H = \int\!\prod_{\vx}d\Lambda_{\vx}\prod_{x}\prod_{k=1}^3 dU_{\vx k}(x_4)
\prod_x d\opsi_x d\psi_x ~\exp\left\{ - S_W(U;\psi;\opsi)
\right\} ,
\en

\noi where $~S_W~$ is the Wilson action with $~U_{\vx 4}(x_4) = 1~$ and
boundary conditions

\eqa
\psi_{\vx}(x_4=N_4a) &=& -\psi_{\vx}^{\Lambda}(x_4=0)~;
\qquad
\opsi_{\vx}(x_4=N_4a) = -\opsi_{\vx}^{\Lambda}(x_4=0)~;
\nonumber \\
\\
U_{\vx k}(x_4=N_4a) &=& U_{\vx k}^{\Lambda}(x_4=0)~.
\nonumber
\ena

\noi The final step is a change of variables

\eq
\psi_{\vx}(x_4=N_4a)\lra \Lambda_{\vx}\psi_{\vx}(x_4=N_4a)~;
\quad
\opsi_{\vx}(x_4=N_4a)\lra \opsi_{\vx}(x_4=N_4a)\Lambda_{\vx}^{\dagger}~,
\en

\noi and similar for the gauge fields $U$. One obtains

\eq
Z_H = \int\!\prod_{\vx}dU_{\vx 4}(x_4=N_4a-a)\prod_{x}\prod_{k=1}^3
dU_{\vx k}(x_4)[d\opsi d\psi] ~\exp\left\{ -S_W(U;\psi;\opsi)\right\}~,
\en

\noi where

\eq
U_{x4} = 1 ~~~\mbox{at}~~~ 0\le x_4 \le N_4a-2a~;
\qquad
U_{\vx 4}(x_4=N_4a-a) = \Lambda_{\vx}~,
             \label{temp_gauge}
\en

\noi and boundary conditions

\eqa
\psi_{\vx}(x_4=N_4a) &=& -\psi_{\vx}(x_4=0)~;
\qquad
\opsi_{\vx}(x_4=N_4a) = -\opsi_{\vx}(x_4=0)~;
\nonumber \\
            \label{antiper}
\\
U_{\vx k}(x_4=N_4a) &=& U_{\vx k}(x_4=0)~.
\nonumber
\ena

\noi Evidently, $Z_H$ coincides with $Z_W$ in the temporal gauge
defined in eq.~(\ref{temp_gauge}). This proves the equivalence of
two approaches.

Two comments are in order.

\begin{itemize}

\item[{\bf i)}] In the functional integral formalism Grassmannian
variables $\eta,\oeta,\zeta,\ozeta$ correspond to the operators
$c,c^{\dagger},d,d^{\dagger}$. However, new Grassmannian variables
$\psi,\opsi$ are connected with $~\eta,\oeta,\zeta,\ozeta~$ in a rather
nontrivial way given in eq.~(\ref{new_grass}). This observation will
appear to be important for the calculation of fermionic matrix
elements.

\item[{\bf ii)}] Antiperiodic boundary conditions in eq.~(\ref{antiper})
stem from the choice of the fermionic Fock space which is, in fact, a
physical assumption. Another choice of the fermionic Fock space gives
another boundary conditions for fermionic variables $\psi,\opsi$ along
the time direction. In details this question will be discussed later.

\end{itemize}

\section{Fermionic currents}
\setcounter{equation}{0}

\subsection{Pseudoscalar current}

Pseudoscalar current $~\hatJ_{\vx}^{(P)}~$ is given by

\eq
\hatJ_{\vx}^{(P)} = ~ :\! i\chi^{\dagger}_{\vx}\gamma_4\gamma_5
\chi_{\vx}\! :
~=~ i(c^{\dagger}_{\vx}d^{\dagger\, T}_{\vx} - d_{\vx}^Tc_{\vx})~,
\en

\noi where $\gamma_4,\gamma_5$ are euclidian $~\gamma$--matrices

\eq
\gamma_{k}
= \left(
\begin{array}{cc}
       0              &  i\sigma_{k}   \\
  -i\sigma_{k}        &       0
\end{array} \right) ;
~~\gamma_{4}
 = \left(
\begin{array}{cc}
       1        &       0         \\
       0        &      -1
\end{array} \right) ;
\quad \quad
\gamma_5 = \gamma_1\gamma_2\gamma_3\gamma_4~.
\en

\noi Evidently, $~\hatJ_{\vx}^{P\,\dagger} = \hatJ_{\vx}^P~$. Let us
derive the functional integral representation of the statistical
average

\eq
\langle \hatJ^{(P)}\rangle
\equiv \frac{1}{Z}\tr \left( \hatJ^{(P)}\,\hatV^{N_4}P_0 \right)~,
\en

\noi where

\eq
\hatJ^{(P)} = \sum_{\vx} \hatJ_{\vx}^{(P)}~.
\en

\noi As well as in the previous section, we assume that the fermionic
Fock space spanned by all possible fermionic states. To derive the
desired expression one should proceed in a same way as outlined in the
second section and use the properties of the Grassmannian coherent
states in eq.~(\ref{properies_coher}). Taking into account

\eq
\Bigl(\oeta_{\vx}\ozeta_{\vx}^T - \zeta_{\vx}^T\eta_{\vx}\Bigr)(x_4)
= \sum_{\vy\vz} \opsi_{\vy}(x_4)\bfB^{1/2}_{\vy\vx}(x_4)\gamma_5
\bfB^{1/2}_{\vx\vz}(x_4) \psi_{\vz}(x_4) ~,
\en

\noi one obtains

\eq
\langle \hatJ^{(P)}\rangle
= -\frac{i}{Z}\int\![dU][d\opsi\psi]~\Bigl( \opsi\bfB
\gamma_5\psi\Bigr)(0) \cdot\exp\{-S_W(U;\opsi;\psi)\}~,
\en

\noi where

\eq
\Bigl(\opsi\bfB\gamma_5\psi\Bigr)(0) = \sum_{\vx\vy}
\opsi_{\vx}(x_4)\bfB_{\vx\vy}(x_4)\gamma_5\psi_{\vy}(x_4)\Big|_{x_4=0}~,
\en

\noi and boundary conditions given in eq.~(\ref{antiper}).

For the zero--momentum pseudoscalar correlator

\eq
\Gamma^{(P)}(\tau) = \frac{1}{Z}\tr \left( \hatV^{N_4-\tau}
\hatJ^{(P)} \hatV^{\tau}\hatJ^{(P)}P_0 \right)~,
\en

\noi one arrives at

\eq
\Gamma^{(P)}(\tau) = -\frac{1}{Z}\int\![dU][d\opsi\psi]~\Bigl( \opsi\bfB
\gamma_5\psi \Bigr)(\tau)\Bigl( \opsi\bfB\gamma_5\psi\Bigr)(0)
\cdot\exp\{-S_W(U;\opsi;\psi)\}~.
\en

\noi One can see that Grassmannian current $~J^{(P)}~$ has no naive
expression
$~J^{(P)}_{naive} = \sum_{\vx} \opsi_x\gamma_5\psi_x~$ .
Instead, it has rather complicated expressions and depends on the
gauge field $U_{x\mu}$. However, in the continuum limit $a\to 0$

\eqa
J^{(P)} &=& \sum_{\vx} \left[ \opsi_{\vx}\gamma_5\psi_{\vx}
- \kappa\sum_{k=1}^3\Bigl( \opsi_{\vx}U_{xk}\gamma_5\psi_{\vx+k}
+ \opsi_{\vx}U_{x-k,k}^{\dagger}\gamma_5\psi_{\vx-k} \Bigr)\right](x_4)
\nonumber \\
\nonumber \\
&\to& (1-6\kappa)\sum_{\vx} \Bigl( \opsi_{\vx}\gamma_5\psi_{\vx}\Bigr)
(x_4) ~\sim~ J^{(P)}_{naive}~.
\ena

\noi  Therefore, at nonzero spacing $~a~$ there is a nonperturbative
renormalization of the matrix element and correlator.

\subsection{Generalized partition function and chemical potential}

Let us derive the functional integral representation for the
generalized partition function $~Z(\lambda_q,\lambda_{\oq})~$ defined
as

\eq
Z(\lambda_q,\lambda_{\oq}) = \tr \left( \hatV^{N_4} \cdot
\exp\Bigl\{ \lambda_q \hatN_q + \lambda_\oq \hatN_\oq\Bigr\}
P_0\right)~,
\en

\noi where

\eq
\hatN_q = \sum_{\vx} c_{\vx}^{\dagger}c_{\vx}~;
\qquad
\hatN_\oq = \sum_{\vx} d_{\vx}^{\dagger}d_{\vx}~.
\en

\noi Choosing the standard fermionic Fock space and repeating
calculations outlined in the second section one obtains

\eq
Z(\lambda_q,\lambda_{\oq}) = \int\![dU][d\opsi d\psi]
~\exp\left\{ -S_W + \delta S_F\right\}~
       \label{Z_lambda_bar}
       \en

\noi with

\eqa
\delta S_F &=& 2\kappa \sum_{\vx}\Bigl[ (e^{\lambda_q}-1)\opsi_{\vx}(a)
P^{(+)}_4 U^{\dagger}_{\vx 4}(0)\psi_{\vx}(0)
+ (e^{\lambda_\oq}-1)\opsi_{\vx}(0)P^{(-)}_4 U_{\vx 4}(0)\psi_{\vx}(a)
\Bigr]
\nonumber \\
\nonumber \\
&& \hspace{1cm}
- (e^{\lambda_q+\lambda_\oq}-1)\sum_{\vx\vy}\opsi_{\vx}(0)P^{(-)}_4
\bfC_{\vx\vy}(0) \psi_{\vx}(0)~,
           \label{delta_S}
\ena

\noi and

\eq
\bfC_{\vx\vy} = \frac{1}{2}\sum_{k=1}^3\Bigl[ \delta_{\vy,\vx+k}U_{xk}
- \delta_{\vy,\vx-k}U_{x-k,k}^{\dagger} \Bigr]\gamma_k~.
\en

The partition function $Z(\mu)$ with nonzero chemical potential $~\mu~$
is given by

\eq
Z(\mu) = \tr\left(\exp\Bigl\{-\frac{1}{T}(\hatH - \mu \hatN)\Bigr\}
P_0\right)
= \tr \left( \hatV^{N_4} \cdot
\exp\Bigl\{ \frac{\mu}{T} \hatN\Bigr\}P_0\right)~,
\en

\noi where $~\hatN = \hatN_q - \hatN_\oq~$. Choosing

\eq
\lambda_q = -\lambda_\oq = \frac{\mu}{T}~.
\en

\noi one obtains

\eq
\delta S_F = 2\kappa \sum_{\vx}\Bigl[ (e^{\mu/T}-1)\opsi_{\vx}(a)
P^{(+)}_4 U^{\dagger}_{\vx 4}(0)\psi_{\vx}(0)
+ (e^{-\mu/T}-1)\opsi_{\vx}(0)P^{(-)}_4 U_{\vx 4}(0)\psi_{\vx}(a)\Bigr].
\en

\noi Making the change of variables

\eq
\psi_{\vx}(x_4) \to  \left\{
\begin{array}{rl}
  e^{ -x_4\mu} \psi_{\vx}(x_4) & ~x_4\ne 0   \\ \\
  e^{ -L_4\mu} \psi_{\vx}(x_4) & ~x_4 = 0
\end{array} \right.
~~\mbox{and}~~~~
\opsi_{\vx}(x_4) \to  \left\{
\begin{array}{rl}
  e^{ x_4\mu} \opsi_{\vx}(x_4) & ~x_4\ne 0   \\ \\
  e^{ L_4\mu} \opsi_{\vx}(x_4) & ~x_4 = 0
\end{array} \right. , ~~~
\nonumber
\en

\noi one obtains the modified fermionic matrix $~\cam(U;\mu)~$

\eqa
\cam_{xy}(U;\mu) &=& \delta_{xy} - 2\kappa
\sum_{k=1}^3\Bigl[ \delta_{y, x+k}P^{(-)}_k U_{xk}
+ \delta_{y,x-k}P^{(+)}_k U_{x-k,k}^{\dagger} \Bigr]
\nonumber \\
\nonumber \\
&& - 2\kappa
\Bigl[ e^{-a\mu}\cdot \delta_{y;x+{\hat 4}}P^{(-)}_4 U_{x4} + e^{a\mu}
\cdot\delta_{y;x-{\hat 4}} P^{(+)}_4 U_{x-{\hat 4};4}^{\dagger} \Bigr]~.
\ena

\noi Evidently, $~\cam(U;\mu)~$ coinsides with the fermionic matrix
for the nonzero chemical potential proposed in \cite{haka}.

\subsection{Scalar `condensate'}

Let us obtain the functional integral representation of the statistical
average $~\langle \hatJ^{(S)}\rangle~$ of the scalar current

\eq
\hatJ^{(S)} = \sum_{\vx} :\!\chi^{\dagger}_{\vx}\gamma_4\chi_{\vx}\! :
~=~ \hatN_q + \hatN_\oq~.
\en

\noi In what follows this average will be refered to as scalar
condensate. Evidently,

\eq
\langle \hatJ^{(S)}\rangle = \frac{1}{Z}\tr\left(\hatJ^{(S)}\,\hatV^{N_4}
P_0 \right)
~=~ \frac{1}{Z}\frac{d}{d\lambda}Z(\lambda)\Big|_{\lambda=0}~,
\en

\noi where

\eq
Z(\lambda) = \tr \left( \hatV^{N_4} \cdot
\exp\Bigl\{ \lambda(\hatN_q + \hatN_\oq)\Bigr\}
P_0\right)~.
\en

\noi From eq.'s~(\ref{Z_lambda_bar}),(\ref{delta_S}) one derives
at $~\lambda_q = \lambda_\oq \equiv \lambda~$

\eqa
\lefteqn{
Z(\lambda) = \int\![dU][d\opsi d\psi]
~\exp\left\{ -S_W + \delta S_F\right\}~;    }~~~~~
\\
\nonumber \\
\delta S_F &=& 2\kappa \Bigl[ (e^{\lambda}-1)\opsi(a)
P^{(+)}_4 U^{\dagger}_{4}(0)\psi(0)
+ (e^{\lambda}-1)\opsi(0)P^{(-)}_4 U_{4}(0)\psi(a)
\nonumber \\
\nonumber \\
&& \hspace{1cm}
- (e^{2\lambda}-1)\opsi(0)P^{(-)}_4\bfC(0)
\psi(0)\Bigr]~,
\nonumber
\ena

\noi where the notations are used

\eq
\opsi(a)P^{(+)}_4 U^{\dagger}_4(0)\psi(0) =
\sum_{\vx}\opsi_{\vx}(x_4=a)P^{(+)}_4
U^{\dagger}_{\vx;4}(x_4=0)\psi_{\vx}(x_4=0)~,
\en

\noi etc.. Finally, one obtains

\eqa
\lefteqn{
\langle \hatJ^{(S)}\rangle = \frac{1}{Z} \int\![dU][d\opsi d\psi]
~e^{ -S_W(U,\psi,\opsi)}         }
          \label{scalar_cond}
\\
\nonumber \\
&& \cdot 2\kappa\Bigl[ \opsi(a)P^{(+)}_4 U^{\dagger}_4(0)\psi(0)
+ \opsi(0)P^{(-)}_4 U_4(0)\psi(a)
- 2\opsi(0)P^{(-)}_4\bfC(0)\psi(0)\Bigr] ~,~~~~~
\nonumber
\ena

\noi Evidently, the Grassmannian current

\eq
J^{(S)} = 2\kappa\Bigl[ \opsi(a)P^{(+)}_4 U^{\dagger}_4(0)\psi(0)
+ \opsi(0)P^{(-)}_4 U_4(0)\psi(a)
- 2\opsi(0)P^{(-)}_4\bfC(0)\psi(0)\Bigr] ~,
\en

\noi does not coincide with

\eq
J^{(S)}_{naive} = \Bigl(\opsi\psi\Bigr)(0) =
\sum_{\vx}\opsi_{\vx}(x_4=0)\psi_{\vx}(x_4=0)~.
\en

\noi  In the continuum limit $a\to 0$ one obtains

\eq
J^{(S)} \to 2\kappa\Bigl(\opsi\psi\Bigr)(0) \sim J^{(S)}_{naive}~.
\en

\noi As well as in the case of the pseudoscalar current, there is a
nonperturbative renormalization of the scalar condensate at nonzero
spacing $~a~$.

\section{Fermionic Fock space and boundary conditions}
\setcounter{equation}{0}

One or another choice of the fermionic Fock space depends on the model
(physical) assumptions. For example, QCD vacuum is expected to have an
equal number of quarks and antiquarks and it puts a restriction on the
choice of the states. The important observation is that boundary
conditions for Grassmannian variables $\psi,\opsi$ along the imaginary
time direction depend on this choice.

Till now it has been assumed that the fermionic Fock space is spanned
by all possible fermionic states, i.e. states

\eq
|\{n_i\};\{m_j\}\rangle = \prod_{i=1}^N (c_i^{\dagger} )^{n_i}
\prod_{j=1}^N (d_j^{\dagger} )^{m_j}|0\rangle~,
              \label{fock1}
\en

\noi where $~n_i,m_j=0,1~$ and $~N=2N_cV_3~$ is a maximal value of
parameters $~i,j~$. This choice of the Fock space results in the
antiperiodic boundary conditions for fermionic Grassmannian variables
in eq.~(\ref{antiper}). Indeed, the trace of any fermionic operator
$~\hatO~$ can be written as

\eqa
\tr_{\chi} \hatO &=&
\int\! [d\oeta d\eta][d\ozeta d\zeta]
[d\oeta^{\prime} d\eta^{\prime}][d\ozeta^{\prime} d\zeta^{\prime}]
% \int_{\eta\zeta}\int_{\eta^{\prime}\zeta^{\prime}}
~ \exp\Bigl\{-\oeta\eta -\ozeta\zeta -\oeta^{\prime}\eta^{\prime}
-\ozeta^{\prime}\zeta^{\prime} \Bigr\}
\nonumber \\
\nonumber \\
&& \cdot \langle\eta\zeta|\hatO|\eta^{\prime}\zeta^{\prime}\rangle \cdot
\tr \Bigl( |\eta\zeta\rangle\langle\eta^{\prime}\zeta^{\prime}|\Bigr)~.
\ena

\noi Choosing Fock space as in eq.~(\ref{fock1}) one obtains

\eq
\tr \Bigl( |\eta\zeta\rangle\langle\eta^{\prime}\zeta^{\prime}|\Bigr)
= \exp\left\{ -\sum_{i=1}^N \Bigl( \oeta_i^{\prime}\eta_i
+ \ozeta_i^{\prime}\zeta_i\Bigr)\right\}~,
\en

\noi and

\eq
\tr_{\chi} \hatO = \int\! [d\oeta d\eta][d\ozeta d\zeta]
~ \exp\Bigl\{ -\oeta\eta -\ozeta\zeta\Bigr\}
\cdot \langle\eta\zeta|\hatO|-\eta;-\zeta\rangle~.
\en

\noi Minuses in  $~|-\eta;-\zeta\rangle~$ presume antiperiodic
boundary conditions for $\psi,\opsi$ along the imaginary time
direction.

However, the choice of the fermionic Fock space made in
eq.~(\ref{fock1}) is not a unique one. As an example, let us consider
the zero temperatures limit $T\to 0$. In this case the main
contribution to the partition function $~Z~$ is expected to come from
the vacuum eigenstate $~|\mbox{vac}\rangle$ :

\eq
Z = \langle\mbox{vac}|\hatV^{N_4} |\mbox{vac}\rangle~;
\qquad T\to 0~.
\en

\noi The corresponding Fock space is supposed to have the equal number
of quarks and antiquarks. Therefore, it looks reasonable to choose the
fermionic Fock space spanned by the vectors

\eq
|\{n_i\};\{m_j\}\rangle = \prod_{i=1}^N (c_i^{\dagger} )^{n_i}
\prod_{j=1}^N (d_j^{\dagger} )^{m_i}|0\rangle
\qquad \mbox{with} \qquad
\sum_i n_i =\sum_i m_i~.
             \label{fock2}
\en

\noi Using the following representation for the delta--symbol

\eq
\delta_{\sum_j n_j ;\sum_j m_j} = \frac{1}{N}\sum_{k=0}^{N-1}
\exp\left\{ \frac{i2\pi}{N}k\sum_j (n_j - m_j)\right\}~,
\en

\noi one obtains

\eq
\tr \Bigl( |\eta\zeta\rangle\langle\eta^{\prime}\zeta^{\prime}|\Bigr)
= \frac{1}{N}\sum_{k=0}^{N-1}  \exp\left\{
-q_k\oeta^{\prime}\eta- q_k^{\ast}\ozeta^{\prime}\zeta\right\} ~,
\en

\noi where

\eq
q_k = \exp\left\{ \frac{2\pi ik}{N}\right\}~;
\qquad k=0;1;\ldots;N-1~.
\en

\noi Therefore,

\eq
\tr_{\chi} \hatO = \frac{1}{N}\sum_{k=0}^{N-1}\int\! [d\oeta d\eta]
[d\ozeta d\zeta] ~ \exp\Bigl\{ -\oeta\eta -\ozeta\zeta\Bigr\}
\cdot \langle\eta\zeta|\hatO|-q_k\eta;-q_k^{\ast}\zeta\rangle~.
\en

\noi In the infinite volume limit $V_3\to\infty$ one arrives at

\eq
\tr_{\chi} \hatO = \frac{1}{2\pi}\int_0^{2\pi}\! d\varphi
\int\! [d\oeta d\eta][d\ozeta d\zeta] ~ \exp\Bigl\{ -\oeta\eta
-\ozeta\zeta\Bigr\} \cdot \langle\eta\zeta|\hatO
|-e^{i\varphi}\eta;-e^{-i\varphi}\zeta\rangle~.
\en

\noi Finally, fermionic boundary conditions for $\psi,\opsi$ are

\eq
\psi_{\vx}(x_4=N_4a) = -e^{i\varphi}\psi_{\vx}(x_4=0)~;
\qquad
\opsi_{\vx}(x_4=N_4a) = -e^{-i\varphi}\opsi_{\vx}(x_4=0)~
               \label{bound}
\en

\noi with integration over $\varphi$ between $0$ and $2\pi$. On the
finite lattice the temperature is nonzero and higher exitation states
can give the (artifact) contributions. Therefore, one may expect that
at $N_4<\infty$ the boundary conditions given in eq.~(\ref{bound})
could be a better choice for the zero temperature calculations (e.g.,
for the hadron spectroscopy study) as compared to the choice of
antiperiodic boundary conditions.

Another interesting case is the finite temperature transition(s) in the
Universe. The baryon asymmetry $~\Delta B~$ of the Universe  is small
and it is expected to be zero in QCD. Therefore, also in this case the
choice of the boundary conditions given in eq.~(\ref{bound}) looks
preferable. It is worthwhile to note that for Polyakov loop $\capbig$
one obtains

\eq
\langle \capbig\rangle = 0~,
\en

\noi and $\langle |\capbig|\rangle $ is expected to be a good order
parameter as in quenched QCD.

Of course, different physical problems need different assumptions about
the structure of the fermionic Fock space. For example, finite
temperature phase transition with nonzero baryon density ($\Delta B\ne
0$) presume another choice of the fermionic Fock space. One can expect
that the phase transition temperature  $T_c$ depends on the choice of
this Fock space and, therefore, on the choice of the boundary
conditions along the forth direction.

\section{Conclusions}\setcounter{equation}{0}

The first goal of this work is  the study of the connection between the
fermionic currents $~\hatJ~$ in the canonical quantization approach and
corresponding currents $J$ in the functional integral (Wilson's)
scheme. In the canonical quantization approach the fermionic currents
$~\hatJ~$ (e.g. pseudoscalar current, etc.) are defined as normal
products of operators $\chi_{\vx}$, $\chi^{\dagger}_{\vx}$, while in
the functional integral formalism the fermionic currents $J$ are
constructed out of Grassmannian variables $\psi_x,\opsi_x$.

As an example, two operator currents $~\hatJ^{(S)} =
~:\!\chi^{\dagger}_{\vx}\gamma_4\chi_{\vx}\! :~$ and $~\hatJ^{(P)} =
~:\! \chi^{\dagger}_{\vx}\gamma_4\gamma_5\chi_{\vx}\! :~$ have been
considered. It appears that corresponding Grassmannian currents $J$
have no naive expressions $~J^{(S)}_{naive}=\opsi_x\psi_x~$ and
$~J^{(P)}_{naive}=\opsi_x\gamma_5\psi_x~$. Instead,
Grassmannian currents have rather complicated expressions and depend on
the gauge field $U_{x\mu}$. However, Grassmannian currents
$~J^{(S,P)}~$ become $~\sim J^{(S,P)}_{naive}~$ in the continuum limit
$~a\to 0~$. Therefore, at nonzero spacing there is a nonperturbative
renormalization of the matrix elements and correlators.

Another scope of this paper is the study of the boundary conditions
along the imaginary time direction.  The important observation is that
boundary conditions for $\psi,\opsi$ along the imaginary time direction
depend on the choice of the fermionic Fock space, and the choice of the
fermionic Fock space depends on the model (physical) assumptions. For
example, in the zero temperatures limit the main contribution to the
partition function $~Z~$ is expected to come from the vacuum eigenstate
$|\mbox{vac}\rangle$ which is supposed to have an equal number of
quarks and antiquarks (if any). In this case it is reasonable to choose
the fermionic Fock space as shown in eq.~(\ref{fock2}) which results in
the boundary conditions for $~\psi,\opsi~$ given in eq.~(\ref{bound}).
One may expect that on the finite lattice with $N_4<\infty$ the choice
in eq.~(\ref{bound}) can be better one for the zero temperature
calculations as compared to the case of antiperiodic boundary
conditions. These boundary conditions look also reasonable for the
study of the finite temperature transition on the cosmological scale
with zero baryon density.

In this paper the derivation of Grassmannian currents has been performed
in the theory with Wilson fermions. It could be interesting to make a
comparison with another versions of lattice QCD, e.g., with staggered
fermions \cite{kosu,stag}.

\section*{Acknowledgements}

This work has been supported by the grant INTAS--00--00111 and RFBR
grant 02--02--17308.

\end{document}